\documentclass[preprint]{rsl}

\title{Wavelet Scattering Networks for \\  Identifying Radio Galaxy Morphologies}
\author{Emma Tolley}

\begin{document}

\maketitle

%
%

\begin{abstract}

Classifying the morphologies of radio galaxies is important to understand their physical properties and evolutionary histories.
A galaxy’s morphology is often determined by visual inspection, but as survey size increases robust automated techniques will be needed. Deep neural networks are an attractive method for automated classification, but have many free parameters and therefore require extensive training data and are subject to overfitting and generalization issues. We explore hybrid classification methods using the scattering transform, the recursive wavelet decomposition of an input image.
We analyse the performance of the scattering transform for the Fanaroff-Riley classification of radio galaxies with respect to CNNs and other machine learning algorithms. We test the robustness of the different classification methods with training data truncation and noise injection, and find that the scattering transform can offer competitive performance with the most accurate CNNs.
\end{abstract}

\section{Introduction}

New facilities for radio interferometry such as the Low-Frequency Array (LOFAR) \cite{lofar},
the Murchison Widefield Array (MWA) \cite{mwa}, the MeerKAT telescope~\cite{meerkat}, and the Australian SKA Pathfinder (ASKAP) telescope~\cite{askap} produce PB of data per year. This data flow rate will only increase;  the next-generation Square Kilometre Array Observatory (SKAO) is expected to produce 600 PB of calibrated data products per year~\cite{skao}.
In this era of data-intensive astronomy, advances in statistics, computer science, and machine learning are beginning to be used in investigations of astronomical data ~\cite{new1, new2}. 

Deep neural networks (DNNs) have been extensively used for a wide range of classification tasks in astronomy, including classifying radio galaxy morphologies~\cite{oldradio, fr23, lisa, fr21,firstbench,mirabestbench,cosfire,brand}.
However, DNNs face problems in \emph{generalizability} (the ability of the network to generalize to new datasets or domains) and \emph{interpretability} (the ability for users to understand the decisions of the network)~\cite{trust1, trust2, gen}.
DNNs must be trained on large datasets that represent the entire domain of each class of objects.
However, radio astronomy datasets are often small, with only a few examples of morphologically diverse objects. Regularization techniques such as data augmentation can help with this issue, but out-of-domain generalizability~\cite{ood} remains an unsolved problem.

It may also be possible to improve  generalizabiltiy and interpretability by hard-coding feature extraction techniques into the network structure, minimising the amount of trainable network parameters. This has been explored for radio galaxy data with group-equivariant CNNs~\cite{fr21}, attention-gating~\cite{mirabestbench}, COSFIRE filters~\cite{cosfire}, and rotational standardization\cite{brand}.
Of particular interest is the scattering transform~\cite{scat} which embeds wavelet transforms into a recursive structure similar to a convolutional neural network.
These wavelet scattering networks are computationally efficient~\cite{scatimage} have been demonstrated to be competitive with traditional neural networks for measuring cosmological parameters~\cite{scatcos1}, but have not yet been evaluated for radio galaxy classification.


\subsection{The Scattering Transform}

\vspace{-0.4em}
The scattering transform is the convolution of an input image $I(x)$ by a series of rotated and dilated wavelet filters. For a complete discussion of the scattering transform and its properties, we refer the reader to ~\cite{scat, scatcon, scaspect, scaphys, scatscos2}.

In this study we construct the scattering transform using the Morlet wavelet $\psi(x) \in \mathbb{R}^2$, which is the product of a Gaussian envelope with a sinusoidal wave.
We define filters  $\psi_{j,l}(x)$ by rotating $\psi(x)$  with
$L$ rotations $r_l = 2 \pi l / L$ in $\mathbb{R}^2$ and by dilating it with $J$ scales $2^j > 1$:
\begin{equation}
\psi_{j,l} (x) = 2^{-2j} \psi(2^{-j}r_l^{-1}x )~.
\end{equation}
Each wavelet filter measures variations of scale $2^j$ in the direction given by $r_l$.
The wavelet transform is  stable and invertible if $\psi_{j,l}$ satisfies a
Littlewood-Paley condition, which requires an additional convolution with a low-pass filter $\phi(x) \in \mathbb{R}^2$, for which we use a 2D Gaussian function.
The scattering transform coefficients are then constructed by convolving the wavelet filters with the input image $I(x)$. To second order, the coefficients are:
\begin{align}
    S_0(x) &= (I \star \phi)(x) \\ 
    S_1(j_1,l_1)(x) & = \bigl( |I\star \psi_{j_1,l_1}| \star \phi \bigr) (x) \\
    S_2(j_1,l_1,j_2,l_2)(x) &= \bigl( \bigl||I\star \psi_{j_1,l_1}| \star \psi_{j_2,l_2} \bigr| \star \phi \bigr )(x)
\end{align}
These scattering transform coefficients provide  a hierarchical multiscale approach to examining $I(x)$ by progressively identifying
 the couplings between structures at different scales.
In the context of neural networks, the wavelet transform acts like a convolutional neural network where the filters are pre-defined.

\section{Radio Galaxy Classification Scheme}

\begin{table}[b]
\small
  \caption{Class sizes of the radio galaxy datasets.}
  \label{tab:data}
  \begin{center}
    \begin{tabular}{ccccc}
      Class & \# train & \# augmented train & \# val. & \# test \\
      \hline \hline
      \vspace{-1em} \\
       \multicolumn{5}{c}{FIRST Dataset} \\
       \hline
      FRI & 395 & 3160 & 50 & 50 \\
      FRII & 824  & 6592 & 50 & 50\\
      Compact & 291 &  2328 & 50  & 50\\
      Bent & 248  & 1984 & 50 & 50\\ 
      \hline
      \vspace{-1em} \\
       \multicolumn{5}{c}{MiraBest Dataset} \\
       \hline
       FRI & 298  & 2384 & 50 & 49 \\
      FRII & 327 & 2616 & 54 & 55\\
    \end{tabular}
  \end{center}
\end{table}

We use the Fanaroff-Riley classification of radio galaxies~\cite{fanaroff} to evaluate the performance of the scattering transform.
This binary classification scheme distinguishes radio galaxies with active nuclei based on their radio luminosity: \emph{FRI} galaxies have a core-brightened structure with steep spectra, whereas \emph{FRII} galaxies are edge-brightened, with bright hotspots at the ends of their lobes. These two different morphologies arise from jet interactions with surrounding environments of different densities. While there are also proposals for more semantic taxonomy of radio galaxies~\cite{bowles}, the  Fanaroff-Riley classification provides a robust benchmark for evaluating classification methods.


Two different overlapping radio galaxy datasets are used for this study.
The first dataset of radio galaxies is a collection and combination of several catalogues using the VLA FIRST (Faint Images of the Radio Sky at Twenty-Centimeters) survey~\cite{firstdata}. It contains galaxies identified as FRI, FRII, compact, or bent sources.
For performance, we trim the $300 \times 300$ pixel images to $ 60 \times 60$ pixels.
The second dataset we use is a collection of radio-loud AGN created from NVSS and FIRST surveys called MiraBest~\cite{mirabest}. In this dataset we only use confidently-classified FRI \& FRII galaxies, and we keep the original $150 \times 150$ pixel image size. For both datasets, we reserve approximately 50 instances of each class for separate validation and test datasets. The remaining data are used for training data, which we augment with 90 degree rotations and reflections of input images. The number of samples for each dataset are shown in Table~\ref{tab:data}. 

We also evaluate performance of our classification schemes against the MNIST~\cite{minst} hand-written digits dataset, a common benchmark in machine learning.

\begin{figure}[t]
  \includegraphics[width=\columnwidth]{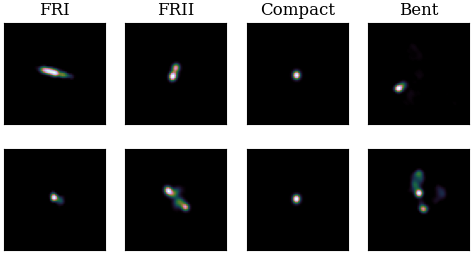}
  \caption{Examples of the four radio galaxy classes from the FIRST dataset.}
  \label{fig:sum}
\end{figure}

\section{Methods}


We use the kymatio library~\cite{kymatio} to construct the scattering transform with Morlet wavelets dilated with $J=3$ spatial scales and $L=8$ angular orientations. We test three different iterations of the scattering transform. First, we use the 2nd-order {\bf scattering transform} including all of the coefficients as defined in Equations 2, 3, and 5. In order to evaluate the effect of the higher order coefficients, we also evaluate a scattering transform that only uses the 0th- and 1st-order coefficients (Equations 2 and 3), and thus only outputs the coefficients of the {\bf wavelet transform}.
Finally, we construct a {\bf reduced scattering transform}
which averages together the different angular modes of the 1st- and 2nd-order coefficients.
For these three schemes the scattering coefficients are input to two different classification methods.

We use linear-kernel support vector classification ({\bf SVC})~\cite{boser, liblinear} using the default \texttt{sklearn} hyperparameters\footnote{scikit-learn.org/stable/modules/generated/sklearn.svm.SVC.html}. SVC is a simple machine learning algorithm which partitions $p$-dimensional input data into different classes with $(p - 1)$ dimensional hyperplanes. SVC can operate directly on the radio galaxy images with no feature extraction,  in which case $p = N_\text{pixel}$, or directly on the scattering transform coefficients.

\begin{table}
\small
  \caption{CNN1 Architecture}
  \label{tab:cnn}
  \begin{center}
    \begin{tabular}{lllc}
      \# & Layer & Description& Activation func.\\
      \hline 
       \hline
      1 & Conv2D & 16 $5 \times 5$ filters & ReLU\\
       2 &  MaxPooling2D & $2\times 2$ pool size \\
       3 &  Conv2D & 16 $5 \times 5$ filters & ReLU\\
     4 & \multicolumn{2}{l}{   GlobalAveragePooling2D}  &  \\
      5 & Dense & 120 nodes & ReLU\\ 
       6 & Dense & 84 nodes & ReLU\\ 
         7 & Dropout & 50\% dropout rate\\ 
      8 & Dense & \multicolumn{2}{l}{ 1/4/10 nodes  \quad\quad sigmoid/softmax}\\ 
      \hline

    \end{tabular}
  \end{center}
\end{table}

We also define a {\bf classifier DNN} network architecture with two hidden layers with 120 \& 84 nodes, rectified linear unit (ReLU) activation functions~\cite{relu}, and 50\% dropout~\cite{dropout} between last hidden layer and output layer.  The final output layer is changed depending on the dataset used. For the multi-class classification with the FIRST (MNIST) dataset, we use four (ten) output nodes with the softmax activation function~\cite{softmax}. For binary classification with the MiraBest dataset, we use a single output node with a sigmoid activation function~\cite{sigmoid}. The network architecture is shown in Lines 5-8 in Table~\ref{tab:cnn}.
The network is trained to classify morphologies using the  scattering transform coefficients as input.

Finally, we develop two complete feature extraction and classification DNN pipelines. The architecture of {\bf CNN1} is inspired by the standard network in~\cite{fr21}.
An overview of the network architecture is shown in Table~\ref{tab:cnn}.
We also define a second neural network called {\bf CNN2} by replacing the global average pooling layer with a second max pooling layer. In total CNN1 has 15,116 free parameters, while CNN2 has 289,676 free parameters.

All of these implementations and the corresponding training scripts are publicly available on GitHub\footnote{https://github.com/epfl-radio-astro/scatternet}.

\section{Results}
We fit each classification method to the training datasets and evaluate performance on the blinded test datasets. The validation dataset is used to evaluate the performance of DNN components during training. We use the Adam optimizer~\cite{adam} with binary cross-entropy loss when training on MiraBest data and categorical cross-entropy loss when training on the FIRST and MNIST datasets. We use sample weights in the loss function to account for class imbalance~\cite{classbalance} in the training datasets. DNNs are fit for 50 epochs with early stopping based on the average of recall obtained on each class (``balanced accuracy'') for the validation dataset.

\begin{figure}
  \includegraphics[width=\columnwidth]{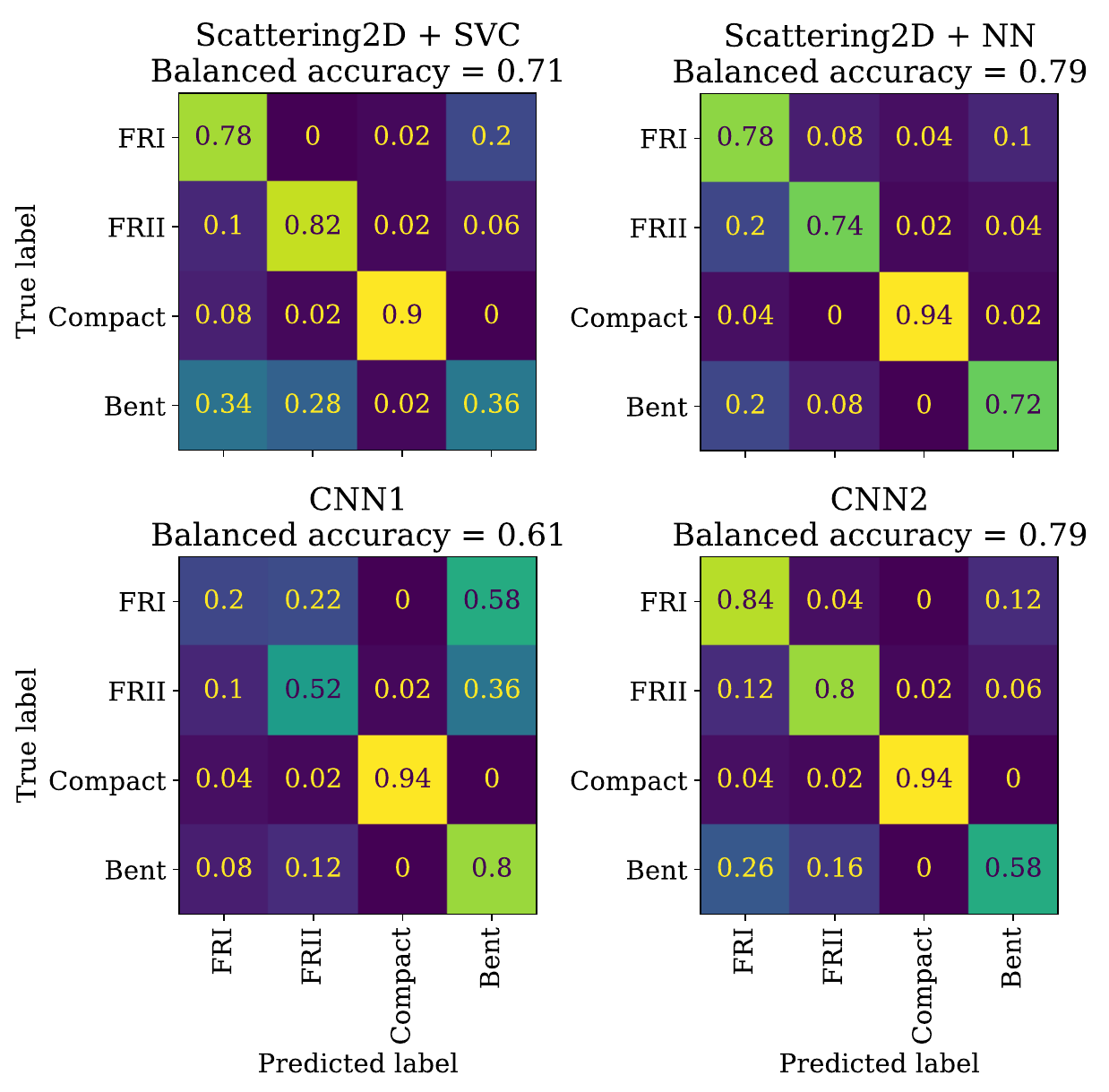}
  \caption{Confusion matrix results for the neural network and 2nd order scattering transform classification methods on the FIRST dataset. The confusion matrix is normalized  over the true conditions. 
  }
  \label{fig:con}
\end{figure}

We show classification performance for the FIRST dataset in Figure~\ref{fig:con} for the CNN1, CNN2, Scattering2D+SVC, and Scattering2D+NN methods. 
Scattering2D+NN and CNN2 show the best performance, with identical balanced accuracy scores.
For the MiraBest dataset, we find that the balanced accuracy of the Scattering2D+NN method is 92\%, slightly better than the CNN2 balanced accuracy of 91\%.

\begin{figure}
  \includegraphics[width=\columnwidth]{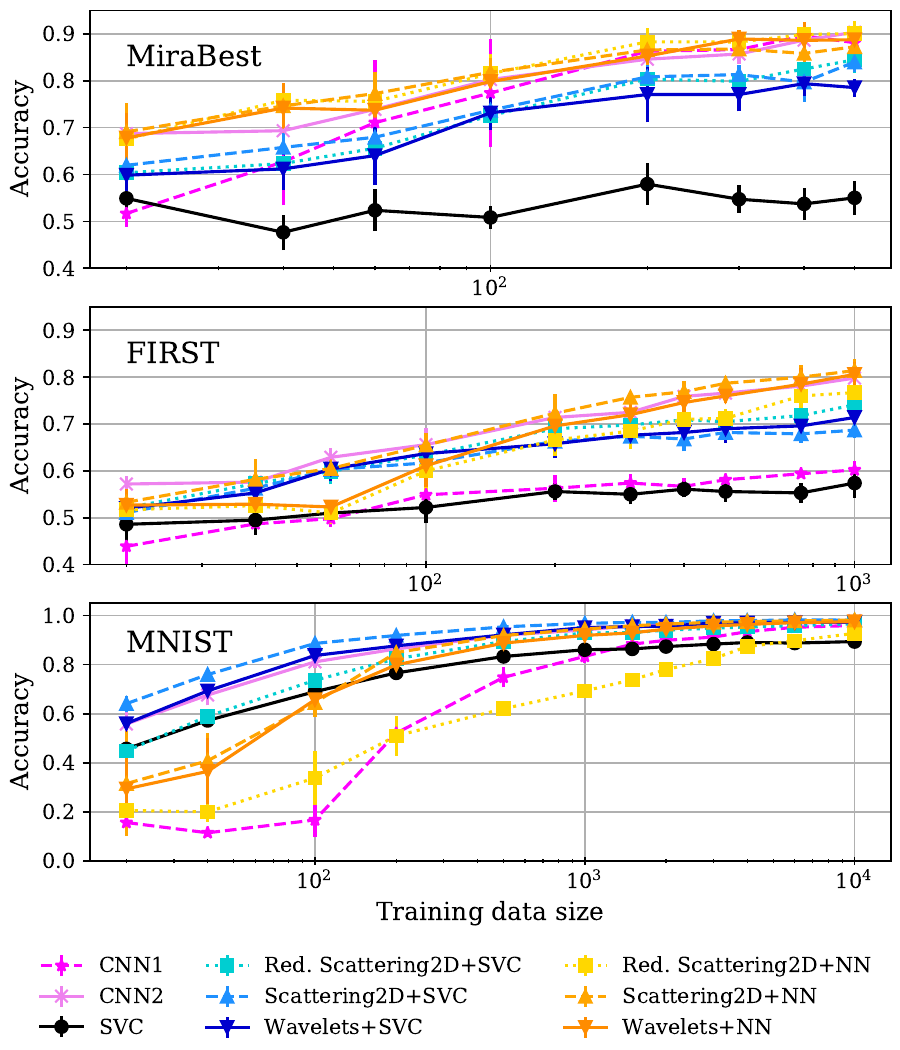}
  \caption{Training data truncation classification results. The $x$-axis shows the total size of the training data (cumulative for all classes) before applying data augmentation. The $y$-axis shows the balanced class accuracy. Error bars represent the standard deviation of five trials.}
  \label{fig:sum}
\end{figure}

However, comparing performance on a single dataset is not enough to demonstrate the robustness of a given classification technique.
We test the generalizability of the different methods 
by {\bf truncating} the full training datasets.
In this test, data augmentation is only applied to the truncated radio galaxy training data \emph{after} the training data have been truncated,  allowing us to evaluate how a network trained on tens or hundreds of training examples might generalize to the test and validation data. We do not apply data augmentation the MNIST data as the classes are not invariant under rotations or reflections.

The results of the truncation tests on the radio galaxy datasets are shown in the top two panels of Figure~\ref{fig:sum}. We see that the solutions which combine the scattering transform with a DNN classifier, and the larger neural network CNN2 consistently perform the best for both radio galaxy datasets. Additionally, we observe that the 2nd order scattering transform consistently outperforms the 1st order scattering transform, demonstrating the added value of using higher-order coefficients.
For the FIRST dataset, the Scattering2D+NN method consistently has the best accuracy.  The smaller network CNN1 struggles with the multi-class classification task, even when trained on the full $\mathcal{O}(1000)$ size training data. 

However, the performance of these different methods is quite different for the MNIST hand-written digits dataset. Here, the 2nd order scattering transform combined with the SVC classifier is consistently the best method, outperforming even CNN2.

\begin{figure}
  \centering
  \includegraphics[width=0.8\columnwidth]{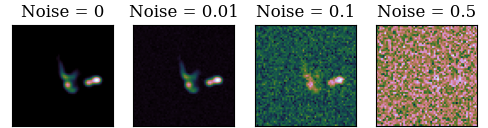}
  \caption{ Example of the noise injection levels for a single FIRST radio galaxy.
  The noise parameter corresponds to the $\sigma$ value of the Gaussian injected noise, relative the the maximum pixel value of the image.
  }
  \label{fig:noise}
\end{figure}

\begin{figure}
  \includegraphics[width=\columnwidth]{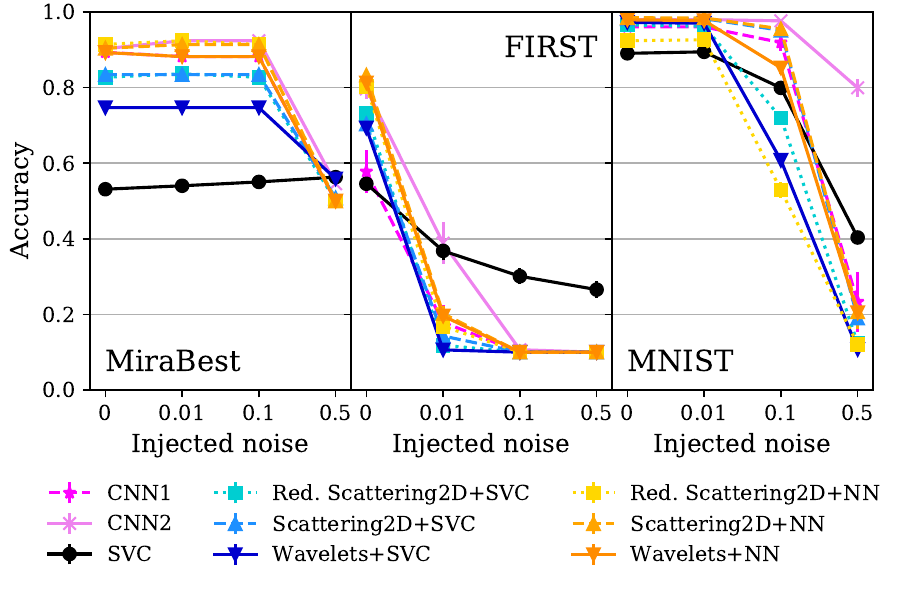}
  \caption{Noise injection test classification results. The $y$-axis shows the balanced class accuracy. }
  \label{fig:noiseres}
\end{figure}

We also test the robustness of the different techniques with {\bf noise injection}. We modify the test data by injecting $\sigma= 1\%$, 10\%, or 50\%  Gaussian noise to each image. An example of the noise injection is shown in Figure~\ref{fig:noise}. Each classification method is fit to the original training data, then evaluated on the different noise-injected test data, with results shown in Figure~\ref{fig:noiseres}. For the MiraBest dataset, again we see that CNN1, CNN2, and the hybrid scattering transform and DNN solutions have similar performance. On the FIRST dataset, none of the classifiers are particularly stable against noise injection, even with $1\%$ injected noise.
Surprisingly, the performance of the simple SVC classifier is the most robust to noise injection across both radio galaxy datasets.
On the MNIST dataset, methods that use DNNs are more robust to noise injection compared to SVC classifiers, an inversion of the truncation test.

\section{Discussion \& Conclusions}

\vspace{-0.4em}
We have shown that the scattering transform can be effectively used for identifying galactic morphology, and is competitive with the best-performing convolutional neural network in this work. We have also shown that including 2nd order coefficients improve feature extraction and classification accuracy.
 We find that our Scattering2D+NN architecture has an accuracy of 79\% on the FIRST dataset, with similar performance to the CNN (79\%) and vision transformer (81\%) classifiers from~\cite{firstbench}. 
On the MiraBest dataset, the Scattering2D+NN architecture has an accuracy of 92\%, comparing favorably to the attention-gated CNN (92\%) in ~\cite{mirabestbench} but not outperforming the standard CNN (94\%) or group-equivariant CNNs from~\cite{fr21}. Using more comprehensive data augmentation as in \cite{mirabestbench,fr21} may improve our results, which seem to be limited by training data size as shown in Figure~\ref{fig:sum}.

We observe strikingly different performance results when evaluating the classification methods on our radio galaxy datasets versus the MNIST dataset. The radio galaxy data propose a much more difficult classification task compared to hand-written digits: galaxy classes are more morphologically diverse, galaxy images are inherently noisier and subject to misclassification error, and contain inherent rotational and translational symmetries that must be learned by any classification method. 

We do not find that the scattering transform improves classifier performance in the data truncation or noise injection generalizability tests. Nevertheless, inclusion of the scattering transform improves network interpretability and drastically reduces the number of trainable parameters, improving the computational and memory requirements. CNN2 and Scattering2D+NN have the same balanced accuracy when evaluated on the FIRST dataset, but Scattering2D+NN uses $\sim 11,000$ parameters whereas CNN2 uses  $\sim 290,000$ parameters. 


We expect that dedicated hyperparameter optimization on each dataset may improve the performance of the classifiers. Additionally, because the scattering coefficients are highly correlated, an additional processing step using principle component analysis to reduce dimensionality may improve performance further.

%
%
\noindent\small
Emma Tolley is with Institute of Physics, Laboratory of Astrophysics, École Polytechnique Fédérale de Lausanne (EPFL), 1290 Sauverny, Switzerland; e-mail: emma.tolley@epfl.ch
\end{document}